\documentclass[conference]{IEEEtran}
\IEEEoverridecommandlockouts

\usepackage{cite}
\usepackage{amsmath,amssymb,amsfonts}
\usepackage{algorithmic}
\usepackage{graphicx}
\usepackage{textcomp}
\usepackage{xcolor}

\usepackage{multirow}
\def\BibTeX{{\rm B\kern-.05em{\sc i\kern-.025em b}\kern-.08em
    T\kern-.1667em\lower.7ex\hbox{E}\kern-.125emX}}
\begin{document}

\title{Rhythmic Foley: A Framework For Seamless Audio-Visual Alignment In Video-to-Audio Synthesis
}

\author{\IEEEauthorblockN{1\textsuperscript{st} Zhiqi Huang$^{\ast}$\thanks{$^{\ast}$  Equal Contribution}$^{\ddagger}$ \thanks{$^{\ddagger}$ Work done during internship at Tencent AI Lab.}}
\IEEEauthorblockA{
\textit{Tsinghua University}\\
Shenzhen, China \\
huangzq22@mails.tsinghua.edu.cn}
\and
\IEEEauthorblockN{1\textsuperscript{st} Dan Luo$^{\ast}$$^{\ddagger}$}
\IEEEauthorblockA{
\textit{Tsinghua University}\\
Shenzhen, China \\
luod23@mails.tsinghua.edu.cn}
\and
\IEEEauthorblockN{2\textsuperscript{nd} Jun Wang$^{\dagger}$\thanks{$^{\dagger}$ Corresponding Author}}
\IEEEauthorblockA{
\textit{Tencent AI Lab} \\
Shenzhen, China \\
joinerwang@tencent.com}
\and 
\IEEEauthorblockN{3\textsuperscript{rd} Huan Liao$^{\ddagger}$}
\IEEEauthorblockA{
\textit{Tsinghua University}\\
Shenzhen, China \\
liaoh22@mails.tsinghua.edu.cn}
\and
\IEEEauthorblockN{4\textsuperscript{th} Zhiheng Li$^{\dagger}$}
 \IEEEauthorblockA{
 \textit{Tsinghua University}\\
 Shenzhen, China \\
 zhhli@mail.tsinghua.edu.cn}
\and
\IEEEauthorblockN{5\textsuperscript{th} Zhiyong Wu$^{\dagger}$}
\IEEEauthorblockA{
\textit{Tsinghua University}\\
Shenzhen, China \\
zywu@sz.tsinghua.edu.cn}
}


\maketitle

\begin{abstract}
Our research introduces an innovative framework for video-to-audio synthesis, which solves the problems of audio-video desynchronization and semantic loss in the audio. By incorporating a semantic alignment adapter and a temporal synchronization adapter, our method significantly improves semantic integrity and the precision of beat point synchronization, particularly in fast-paced action sequences. Utilizing a contrastive audio-visual pre-trained encoder, our model is trained with video and high-quality audio data, improving the quality of the generated audio. This dual-adapter approach empowers users with enhanced control over audio semantics and beat effects, allowing the adjustment of the controller to achieve better results. Extensive experiments substantiate the effectiveness of our framework in achieving seamless audio-visual alignment.
\end{abstract}

\begin{IEEEkeywords}
Video-to-audio Generation, Diffusion, Foley, Audio-visual Alignment
\end{IEEEkeywords}

\section{Introduction}
In the current era of digital content creation, from filmmakers to social video creators, more and more people wish to turn their creative visions into actual outputs. As text-to-video models continue to improve and evolve, the demand for adding vivid and fitting background audio to silent videos is also increasing. To realize their creative ideas, creators need a great deal of control. For social video creators, this control should be simple yet powerful, while for filmmakers, it should be precise and limitless. Our goal is to enable creators to quickly iterate on their ideas and generate satisfactory audio through simple controls.

Currently, there are two main methods for video-to-audio generation, each with its advantages and disadvantages. The first method~\cite{zhang2024foleycrafter,wang2024v2a} builds on text-to-audio generation by adding video event timestamps or video frame features to control the alignment of semantics and sound events. The second method~\cite{luo2024diff,xu2024video} directly trains a video-to-audio diffusion model. Although these methods can achieve video-to-audio generation, they still have three main limitations. \textbf{(i) Asynchronous.} The first method is essentially still text-to-audio generation. Even with the addition of timestamps for alignment control, the instantaneous nature of sound events means that current methods cannot ensure perfect alignment between the generated audio and the actions in the video. \textbf{(ii) Low-quality.} The second method uses the entire video as input, aligning video and audio on the timeline, making more efficient use of the video's temporal and visual information. However, due to the difficulty of obtaining high-quality video-audio datasets, models trained with this method cannot guarantee the generation of high-quality audio. \textbf{(iii) Semantic loss.} Additionally, some details may be lost during video encoding.

To overcome these limitations, we propose a new video-to-audio generation framework that adds two adapters to the pre-trained video-to-audio model: a semantic alignment adapter and a temporal synchronization adapter. Regarding \textbf{semantic loss}, the semantic alignment adapter supplements the semantic information and provides a simple audio event timestamp to control the approximate time of semantic occurrences. The semantic information generated by the large-scale visual understanding model~\cite{li2024mini,chen2024sharegpt4video,zhang2024internlm}, can produce more comprehensive, temporally ordered descriptions compared to the simple captions provided by the dataset. These descriptions include more details and contextual information, helping the model to better understand the content of the video. The timestamp can help the model better align and associate information between video and audio, thereby improving learning effectiveness. 

To solve the problem of \textbf{asynchronous}, the temporal synchronization adapter provides the exact timing of instantaneous audio events, significantly improving the temporal synchronization between audio and video.
For the issue of \textbf{poor audio quality generated}, we use the encoder aligned with audio and video to perform visual sampling on pure audio data. We use high-quality audio data for training the temporal adapter, which greatly enhances the quality of audio generation.

The main contributions are summarized below:\footnote{Our demos are available at  https://angelalilyer.github.io/RhythmicFoley/}
\begin{itemize}
\item[1)] 
The semantic alignment adapter and temporal synchronization adapter can be used together or separately, enabling the generation of audio with complete semantic content and finer temporal alignment. 

\item[2)] 
During our training process, we incorporated a substantial amount of high-quality pure audio data, which significantly improved the quality and robustness of the generated audio. 

\item[3)] 
We propose a framework: Rhythmic Foley, which generates high-quality, fine-grained aligned audio for silent videos. We conduct extensive quantitative and qualitative experiments to verify the effectiveness of our method.

\end{itemize}

\section{RELATED WORK}
\label{sec:format}

\textbf{Text-to-Audio Generation} The traditional Text-to-Audio systems primarily rely on the signal processing methods and machine learning to generate audio, which have limitations in generation quality and computational costs. AudioLDM\cite{liu2023audioldm,liu2024audioldm} improves these aspects by using pre-trained CLAP model\cite{elizalde2023clap} to train latent diffusion models with audio embeddings, conditioned on text embeddings during sampling. Some researchers\cite{wu2023large,yuan2024retrieval} have proposed data augmentation strategies, which have improved the quality and robustness of the audio generation. Other new researches~\cite{vyas2023audiobox,borsos2023audiolm} have also achieved better generation quality and generalization. For example, Tango~\cite{ghosal2023text,majumder2024tango}, outperforms AudioLDM on most metrics. This improvement is attributed to the adoption of audio pressure level-based sound mixing for training set augmentation. Baton~\cite{liao2024baton} and Tango2~\cite{majumder2024tango} curated a preference dataset with human and CLAP annotations respectively, fine-tuning text-to-audio models via reinforcement learning strategies. This achieved better text-audio alignment based on both automatic and manual metrics.

\textbf{Video-to-Audio Generation} In recent years, the field of video-to-audio generation has been extensively explored by researchers, who have developed various innovative approaches. Some methods involve fine-tuning text-to-audio models and integrating additional video information such as visual encodings or timestamps of video events to condition the generation of audio. Examples of such approaches include FoleyCrafter~\cite{zhang2024foleycrafter}, SonicVisionLM~\cite{xie2024sonicvisionlm}, and V2A-Mapper~\cite{wang2024v2a}. Additionally, there are models~\cite{du2023conditional,xu2024video} that have been trained from the ground up specifically for video-to-audio generation using diffusion models. Simian et al. present Diff-Foley~\cite{luo2024diff}, a synchronized Video-to-Audio synthesis method with a latent diffusion model that generates high-quality audio with improved synchronization and audio-visual relevance. Beyond these diffusion models, there are also video-to-audio generation methods~\cite{chen2024semantically,panagopoulou2023x} that are based on large language models. These leverage the semantic consistency between video and audio, allowing for a mutual transmodal conversion.

\section{method}
\label{sec:method}
Providing a silent video, our approach can generate audio that is both semantically aligned and precisely timed to match key moments. Our framework is based on the video-to-audio synthesis, where the audio-visual alignment encoder uses the methodologies and weights from Diff-Foley~\cite{luo2024diff}. For video comprehension, we employ Mini-Gemini~\cite{li2024mini} to extract visual content from the video. This section will delve into the specifics of our model's architecture, along with the procedures for training and inference.
\begin{figure}[htb]
\begin{minipage}[b]{1.0\linewidth}
  \centering
  \centerline{\includegraphics[width=9cm]{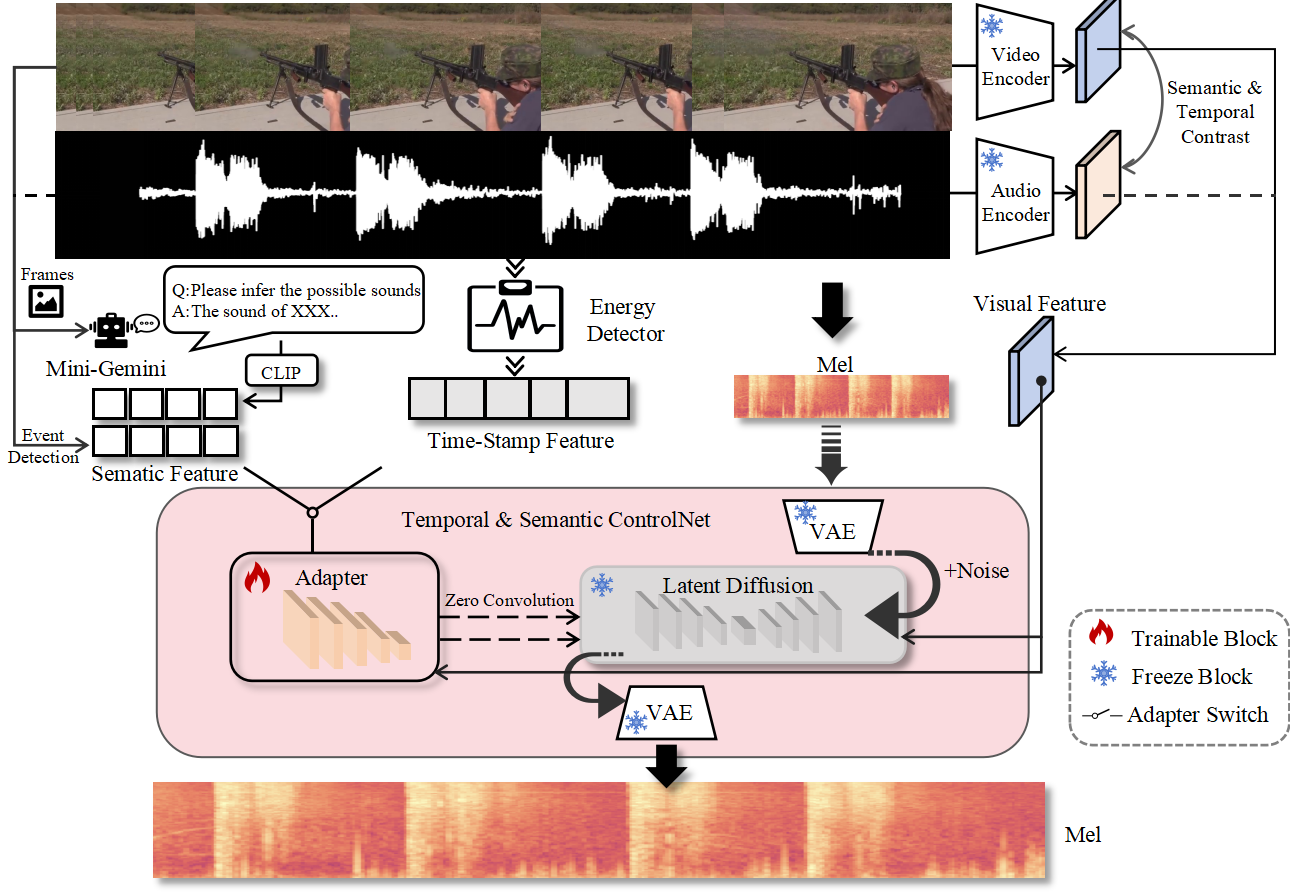}}

\end{minipage}
\caption{An illustration of our framework during the training phase.}
\label{fig:train}
\end{figure}
\begin{figure}[htb]
\begin{minipage}[b]{1.0\linewidth}
  \centering
  \centerline{\includegraphics[width=9cm]{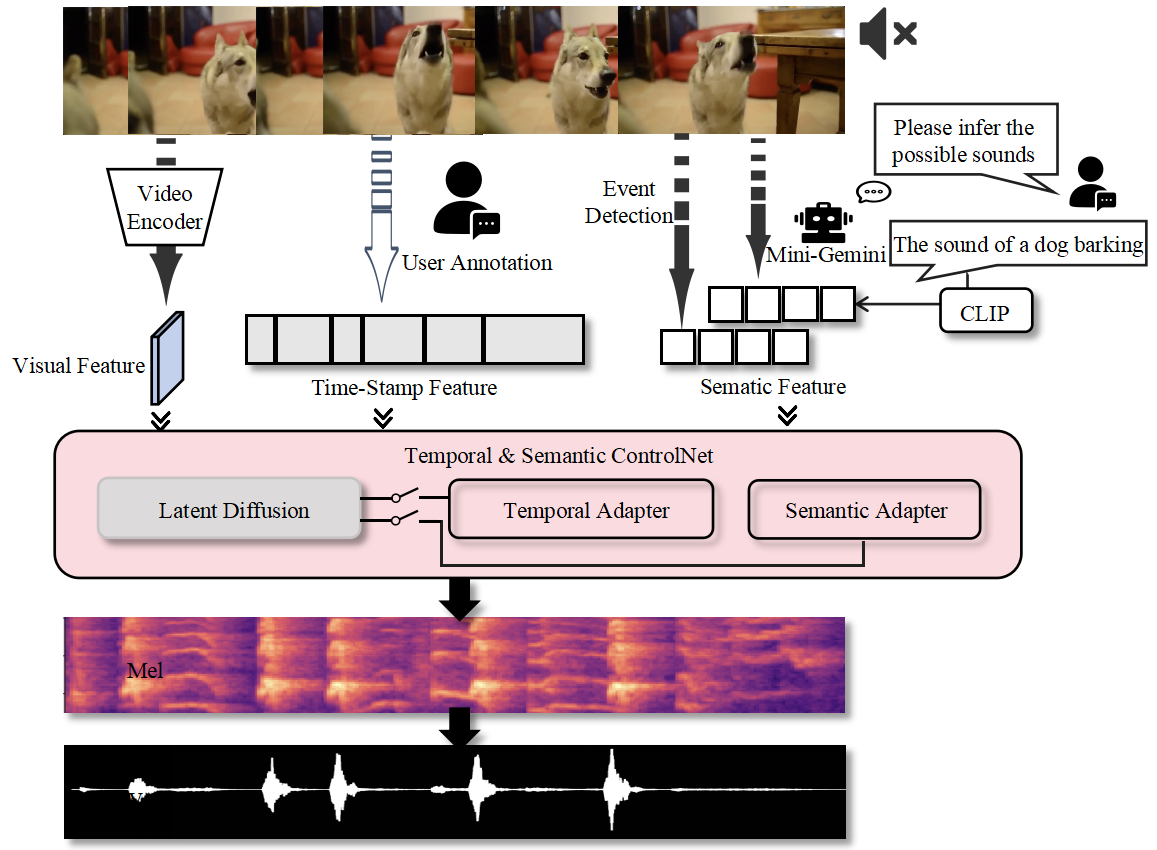}}

\end{minipage}
\caption{An illustration of our framework during the inference phase.}
\label{fig:infer}
\end{figure}

\subsection{Training-time Framework}
As shown in Fig. \ref{fig:train}, we input video or audio data into their respective encoders to acquire aligned features that contain visual and auditory information. The ground truth is also encoded into a mel spectrum for subsequent processing. In addition, we have designed two adapters for different training conditions, which share the same architecture with the UNet encoder~\cite{ronneberger2015u} of the video-to-audio generator, which is the same as ControlNet~\cite{zhang2023adding}. These adapters take the latent embeddings from the UNet and extra conditional embeddings as input and add their outputs as residuals to the UNet's output. 

During the training stage, only the adapter layers are trainable while the UNet blocks are frozen. The loss function used is defined as:

\begin{equation}
L=E_{z_{0},t,c_{t},c_{f},\epsilon\sim N(0,1) }||\epsilon-\epsilon_{\theta }(z_{t},t,c_{t},c_{f})||_{2}^2, 
\label{eq:loss}
\end{equation}
where \(z_{0}\) represents the data in the latent space, \(c_{t}\) and \(c_{f}\) are the video conditions such as semantics and timestamps and the latent audio condition, respectively.

\textbf{The Semantic Alignment Adapter} uses text descriptions extracted by Mini-Gemini~\cite{li2024mini} along with the outputs of event detectors to enhance the semantic completeness and synchronicity of the generated audio. The video event detection is a simple network ~\cite{he2016identity,xie2024sonicvisionlm} trained to identify the occurrence of sound events during a certain period. It helps models pay attention to these visual cues, which is crucial for the transformation of video content into audio. With the significant advancements in large language models~\cite{liu2024st,li2024llava}, there has been substantial progress in multimodal understanding~\cite{zhu2023minigpt,chen2023minigpt,li2024mini}. They can list possible sound events based on visual content, saving the need for manual labeling. During the training of this module, we have employed the VGGSound\cite{chen2020vggsound} and AudioSet\cite{gemmeke2017audio} datasets as our training data. 



\textbf{The Temporal Synchronization Adapter} is designed to synchronize the cadence of audio generation with the motion in the video, employing a timestamp conditional mask to guide the precise timing of sound effect generation. Derived from an energy detector~\cite{mcfee2015librosa}, this mask adeptly pinpoints onset events in the audio stream. For the training of this module, we have collected sound effect data to more accurately learn the rhythm of the audio. The training data for this section was collected from various high-quality audio websites. 

\subsection{Inference-time Framework}
In the inference stage, as shown in \ref{fig:infer}, we have the option to employ the semantic alignment adapter and temporal synchronization adapter either separately or in tandem with combined weighting. For the temporal adapter, if users have higher demands for precise dubbing, we can provide a visualization annotation tool that can achieve better generation results. The semantic adapter, on the other hand, obtains its conditional input through the dialogue analysis of Mini-Gemini~\cite{li2024mini} and detection by visual event detectors. Users can fine-tune the strength of both adapters based on the actual generation results, employing the framework to produce audio that is both rhythmically precise and semantically comprehensive.

\section{Experiments}
\label{sec:experiment}
In this section, we first introduce the experimental setup. Then, we conducted experiments to compare our method with previous approaches across several dimensions: semantic alignment, temporal synchronization, and audio quality. Following this, we evaluate some objective and subjective metrics to demonstrate the effectiveness and superiority of our model.
 The specifics of these experiments and their corresponding analyses will be outlined below.
\subsection{Experimental Setup}

During the training phase, our model was trained for 30 epochs on the VGGSound\cite{chen2020vggsound}, AudioSet\cite{gemmeke2017audio}, and pure audio dataset.
During the evaluation phase, We conducted a comprehensive evaluation of our method, comparing it with FoleyCrafter~\cite{zhang2024foleycrafter}, and techniques mentioned in its paper,including Diff-Foley~\cite{luo2024diff}, V2A-Mapper\cite{wang2024v2a}, and Seeing-and-hearing~\cite{xing2024seeing}. We adhered strictly to the experimental setup outlined in the article for our experiments. These evaluations were performed on the AVSync15\cite{zhang2024audio} and VGGSound\cite{chen2020vggsound} datasets, employing a variety of metrics to assess semantic alignment and audio quality, including Frechet Distance (FID)~\cite{heusel2017gans}, CLIP similarity~\cite{radford2021learning}, and Mean KL Divergence (MKL)~\cite{iashin2021taming}. Furthermore, we utilized onset detection accuracy (Onset Acc) and onset detection average precision (Onset AP)~\cite{du2023conditional} to evaluate the synchronization accuracy of the generated audio.

\subsection{Quantitative and Qualitative Comparison}

\textbf{Quantitative Comparison}
TABLE \ref{tab:matric1} presents a comparative evaluation of our approach against state-of-the-art models in terms of semantic alignment. We assessed three key metrics multiple times on a subset of the VGGSound and AVSync15 test sets, and the table displays the mean values of these tests. In our experiments, Rhythmic-Foley demonstrated superior performance on Frechet Distance and CLIP similarity. Although the improvement in the MKL metric is not as pronounced as in FID and CLIP similarity, the comparable performance indicates that our method maintains a high level of efficacy across different evaluation criteria.
Additionally, TABLE \ref{tab:matric3} illustrates the temporal synchronization outcomes on the AVSync15 dataset. From these evaluations, it is evident that our method excels in both semantic alignment with visual prompts and temporal synchronization. Our approach not only provides a more precise understanding of the audio-visual content but also generates audio that corresponds more accurately with the video. This superior performance highlights the effectiveness of our model in achieving a more coherent and synchronized audio-visual experience, setting a new benchmark in the field.

\begin{table}[t]
\caption{Quantitative results in terms of semantic alignment}

  \label{tab:matric1}
  \centering
  \setlength{\tabcolsep}{5pt} 
  \begin{tabular}{@{}|c|c|c|c|c|@{}}
    \hline
    Testset & Method & FID$\downarrow$ & CLIP$\uparrow$ & MKL$\downarrow$\\
    \hline
    \multirow{4}*{VGGSound} & Diff-Foley~\cite{luo2024diff} & 29.03 &  9.172 &  3.318 \\
    ~ & V2A-Mapper~\cite{wang2024v2a} &  24.16 & 9.720 & 2.654 \\
    ~ & FoleyCrafter~\cite{zhang2024foleycrafter} &  19.67 & 10.70 & 2.561 \\
    ~ & Rhythmic-Foley(ours) &  {\bf15.26} &  {\bf20.82} & \textbf{1.684}\\
    \hline
    \multirow{4}*{AVSync} & Diff-Foley~\cite{luo2024diff} & 65.77 &  10.38 &  1.963 \\
    ~ & Seeing and Hearing~\cite{xing2024seeing} &  65.82 & 2.033 & 2.547 \\
    ~ & FoleyCrafter~\cite{zhang2024foleycrafter} &  36.80 & 11.94 & {\bf1.497} \\
    ~ & Rhythmic-Foley(ours) &  {\bf32.92} &  {\bf18.92} & 1.524 \\
  \hline
  \end{tabular}
\end{table}


  

\begin{table}[t]
  \caption{Quantitative results in terms of temporal synchronization }
  
  \label{tab:matric3}
  \centering
  \setlength{\tabcolsep}{5pt} 
  \begin{tabular}{@{}|c|c|c|c|@{}}
    \hline
    Method &  Onset AP$\uparrow$ & Onset ACC $\uparrow$ \\
    \hline
    Diff-Foley~\cite{luo2024diff} &  66.55 & 21.18 \\
    Seeing and Hearing~\cite{xing2024seeing} &  60.33 & 20.95  \\
    FoleyCrafter~\cite{zhang2024foleycrafter} &  68.14 & 28.48  \\
    Rhythmic-Foley(ours) &  {\bf72.23} &  {\bf 35.68} \\
  \hline
  \end{tabular}
\end{table}

\begin{figure}
    \centering
    \includegraphics[width=1.0\linewidth]{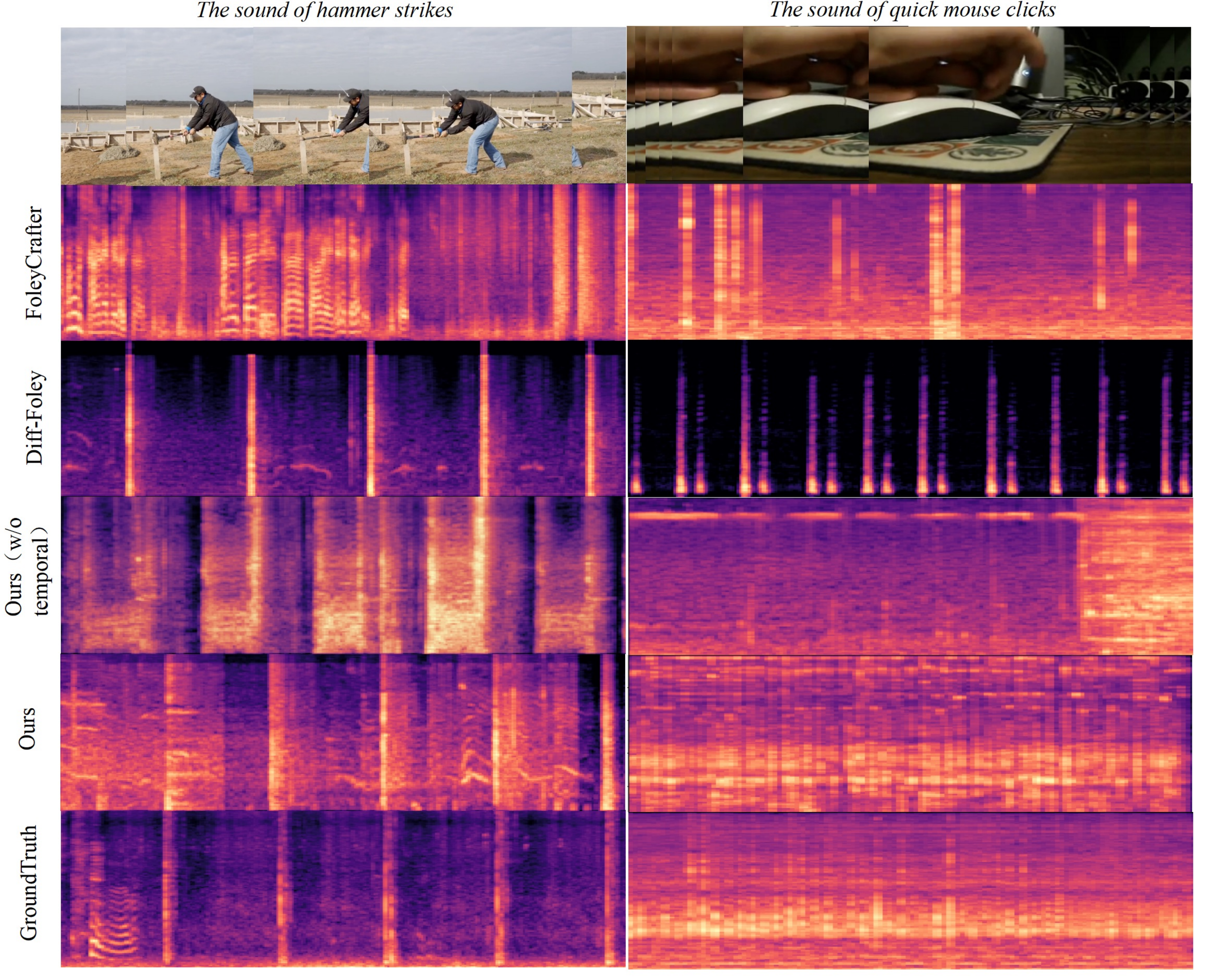}
    \caption{Mel spectrogram comparision}
    \label{fig:contrast_mel}
\end{figure}

\textbf{Qualitative Comparison}
We provide the mel spectrogram of the generated audio and ground in Fig. \ref{fig:contrast_mel} for qualitative comparison. As shown in Fig. \ref{fig:contrast_mel}, our model is able to extract the correct semantics and the precise timing of events from the video. Therefore, it achieves better semantic alignment and higher temporal synchronization compared to other models. Without the temporal adapter, our model can detect events occurring in the video and predict the coarse occurrence and duration of these events through the semantic alignment adapter. The complete Rhythmic Foley, with the help of the temporal synchronization adapter, can also predict more accurate and precise timings of instantaneous actions. Therefore, Rhythmic Foley can achieve SOTA  in these aspects.

\begin{table}[htbp]
  \caption{Average User Ranking (AUR) in terms of audio quality, semantic alignment, and temporal synchronization.}
  \label{tab:matric4}
  \centering
  \setlength{\tabcolsep}{5pt} 
  \begin{tabular}{@{}|c|c|c|c|@{}}
    \hline
    Method & Quality$\uparrow$ & Semantic$\uparrow$ & Temporal$\uparrow$\\
    \hline
    Diff-Foley~\cite{luo2024diff} & 3.283 & 3.467 & 3.260\\
    FoleyCrafter~\cite{zhang2024foleycrafter} & 3.387 & 3.868 & 2.587\\
    Rhythmic-Foley(ours) & \textbf{3.951} & \textbf{3.917} & \textbf{4.236}\\
  \hline
  \end{tabular}
\end{table}
\subsection{User Study} 
We further invited users to evaluate the audio quality, semantic consistency, and temporal alignment of audio generated by different methods. The Average Human Ranking (AHR) was employed to rate each result on a scale of 1 to 5 (with higher scores indicating better performance). Respondents rated a set of audio clips generated by different methods for the same video. We collected 2025 responses, which are presented in TABLE \ref{tab:matric4}. Our method achieved the best in the subjective evaluation across three key aspects.

\subsection{Ablation Study} 
To validate the efficacy of the semantic and temporal adapter in the task of video-audio generation, we conducted a series of ablation studies on the VGGSound and AVSync15 test sets. 

\textbf{Ablation on semantic adapter}
For this module, we conducted experiments using only event detection features, only clip output features, and our method (event + clip) as inputs. The TABLE \ref{tab:matric5} shows that our method performs the best, as the two features complement each other, enabling the model to learn more accurate semantics from the representations.

\begin{table}[htbp]
  \caption{Ablation on semantic adapter tested on the Vggsound}
  \label{tab:matric5}
  \centering
  \setlength{\tabcolsep}{5pt} 
  \begin{tabular}{@{}|c|c|c|c|@{}}
    \hline
    Method & FID$\downarrow$ & CLIP$\uparrow$ & MKL$\downarrow$\\
    \hline
    w/o events  &  15.26 & 20.51 &  1.896 \\
    w/o semantic  &  18.38 & 18.79 &  2.784 \\
    Rhythmic-Foley(ours) &  {\bf15.06} &  {\bf20.82} & \textbf{1.684}\\
  \hline
  \end{tabular}
\end{table}

\textbf{Ablation on temporal adapter}
As illustrated in TABLE \ref{tab:matric6}, the first method employs video features aligned with audio as input features for training, which yields results comparable to our approach (where audio features aligned with video are used for training), albeit slightly inferior, with a negligible difference. However, models lacking a temporal adapter possess only semantic control information without frame-level timestamp details, leading to a marked decrease in the accuracy of time synchronization and the audio quality.
\begin{table}[htbp]
  \caption{Ablation on temporal adapter tested on the AVSync15.}
  \begin{center}
  \label{tab:matric6}
  \centering
  \setlength{\tabcolsep}{5pt} 
  \begin{tabular}{|c|c|c|c|}
    \hline
    Method &  Onset AP$\uparrow$ & Onset ACC $\uparrow$ & VISQOL $\uparrow$ \\
    \hline
    trained by videos &  70.31 & 34.18 & 3.175\\
    w/o temporal &  66.33 & 21.95 & 3.117  \\
    Rhythmic-Foley(ours) &  {\bf72.23} &  {\bf 35.68} & {\bf3.215} \\
  \hline
  \end{tabular}
\end{center}
\end{table}

\section{Conclusion}
\label{sec:conclusion}

In this paper, we introduce a novel framework for video-to-audio generation, which integrates two adapters into the pre-trained video-to-audio generation model: the semantic alignment adapter and the temporal synchronization adapter. These adapters address the issues of semantic loss in audio generation and the asynchrony between audio and video actions, respectively. Furthermore, our training process involves mixing audio and video data, employing an audio encoder that is aligned with the video encoder. Through extensive quantitative experiments and user studies, we have demonstrated the effectiveness of our method, which significantly enhances the quality of video-to-audio generation.

\bibliographystyle{IEEEbib}
\bibliography{strings,refs}

\end{document}